\documentclass[aps, twocolumn, showpacs, letterpaper]{revtex4}

\usepackage{amsmath}
\usepackage{amssymb}
\usepackage{graphicx}
\usepackage{xspace}
\usepackage{upgreek}
\usepackage{accents}

\newcommand{\eg}{{e.g.,\/}\xspace}
\newcommand{\ie}{{i.e.,\/}\xspace}

\newcommand{\gap}{\mbox{}}

\newcommand{\eq}[1]{(\ref{#1})}
\newcommand{\Eq}[1]{Eq.~(\ref{#1})}
\newcommand{\Eqs}[1]{Eqs.~(\ref{#1})}

\newcommand{\Sec}[1]{Sec.~\ref{#1}}

\newcommand{\Fig}[1]{Fig.~\ref{#1}} 
\newcommand{\Ref}[1]{Ref.~\cite{#1}}
\newcommand{\Refs}[1]{Refs.~\cite{#1}}

\newcommand{\mc}[1]{\mathcal{#1}}

\newcommand{\favr}[1]{\langle #1 \rangle}
\renewcommand{\vec}[1]{{\boldsymbol{\rm #1}}}
\newcommand{\oper}[1]{\hat{\vec{#1}}}
\newcommand{\proj}[1]{\bar{#1}}
\newcommand{\uproj}[1]{\underaccent{\bar}{#1}}
\newcommand{\fvec}[1]{\mathsf{#1}}
\newcommand{\foper}[1]{\hat{\fvec{#1}}}
\newcommand{\fvecp}[1]{\proj{\fvec{#1}}}

\newcommand{\pd}{\partial}
\newcommand{\pdf}{\mathit{\pd f}}
\newcommand{\del}{\nabla}
\newcommand{\kpt}[1]{{\kern #1 pt}}
\newcommand{\taufo}{\tau_\text{\tiny FO}}
\newcommand{\grav}{\bar{\text{g}}}
\newcommand{\vecgrav}{\textbf{g}}
\newcommand{\lie}{\pounds}

\begin{document}

\title{Vlasov equation and collisionless hydrodynamics adapted to curved spacetime}

\author{I.~Y. Dodin and N.~J. Fisch}
\affiliation{Department of Astrophysical Sciences, Princeton University, Princeton, New Jersey 08544, USA}
\date{\today}

\begin{abstract}
The modification of the Vlasov equation, in its standard form describing a charged particle distribution in the six-dimensional phase space, is derived explicitly within a formal Hamiltonian approach for arbitrarily curved spacetime. The equation accounts simultaneously for the Lorentz force and the effects of general relativity, with the latter appearing as the gravity force and an additional force due to the extrinsic curvature of spatial hypersurfaces. For an arbitrary spatial metric, the equations of collisionless hydrodynamics are also obtained in the usual three-vector form.
\end{abstract}




\maketitle

\section{Introduction} 
\label{sec:intro}

The Vlasov theory for curved spacetime is well developed, yet has been receiving but scant attention in plasma physics. The reason for this may be that the existing theory relies on covariant formulation \cite{tex:ehlers71, book:cercignani}, which renders difficult to use the intuition available through standard formulations of plasma physics. The so-called $3+1$ formalism could help solve this problem, by projecting the general relativistic (GR) equations on the more conventional three-dimensional (3D) space; specifically, the Maxwell's equations and the particle motion equations can be put in a three-vector form similar to that in the Minkowski metric \cite{ref:thorne82, ref:evans88, arX:gourgoulhon07, ref:jantzen92, ref:bini01, ref:bini97, tex:bini98a, tex:bini98b}. However, a question remains how exactly the Vlasov equation, in its standard form describing a charged particle distribution in the 6D phase space \cite{book:stix}, is modified when spacetime is curved.

To answer this question is the main purpose of the present paper. The same problem was previously addressed \textit{ad hoc} for a specific metric in \Refs{book:bernstein, ref:dettmann93} and more generally in \Refs{ref:debbasch09a, ref:debbasch09b}. However, in the latter case, electromagnetic interactions were not included, and the spacetime basis was not identified that would yield conventional three-vector equations of collisionless plasma hydrodynamics (see below). To close the theory, one thereby needs to specify the $3+1$ equations of charged particle motion explicitly and in the form analogous to that commonly used for the Minkowski metric~\cite{book:stix}. 

The routine approach to these equations that involves the full-fledged machinery of differential geometry \cite{ref:jantzen92, ref:bini97, ref:bini01, tex:bini98a, tex:bini98b} cannot be employed for this purpose. Also, their general form is reported by different authors in different forms which are neither manifestly equivalent, nor always accurate \cite{foot:correct}. Thus, it seems warranted to rederive the corresponding equations from scratch, particularly, in the Hamiltonian representation that is immediately applicable to the Vlasov theory. A spin-off here is that, once the Hamiltonian formalism is developed, the covariant approach used in \Refs{ref:debbasch09a, ref:debbasch09b} becomes unnecessary. Instead of generalizing \Refs{ref:debbasch09a, ref:debbasch09b} then, one can as well make the $3+1$ Vlasov theory self-contained and \textit{reformulate} it in a manner familiar from the standard plasma physics. This constitutes the second purpose of our paper. 

Finally, our third purpose is to derive hydrodynamic equations from the Vlasov theory in curved spacetime in their usual three-vector form. Unlike approaches \textit{postulating} a hydrodynamic closure \cite{ref:dettmann93, ref:gailis94, ref:gailis95, ref:gailis97}, this will yield a fundamental fluid treatment of plasmas in the collisionless limit, where the commonly used ideal-fluid approximation \cite{ref:thorne82, ref:zhang89, ref:gailis95, arX:gourgoulhon07} does not apply.

Specifically, our results can be summarized as follows. The three-dimensional dynamics of a charged particle in an arbitrary spacetime metric, traditionally addressed within differential geometry, is reformulated in terms of linear algebra and Hamiltonian formalism. The modification of the Vlasov equation, in its standard form describing a charged particle distribution in the 6D phase space, is then derived explicitly. The equation accounts simultaneously for the Lorentz force and the effects of general relativity, with the latter appearing as the gravity force and an additional force due to the extrinsic curvature of spatial hypersurfaces. For an arbitrary spatial metric, the equations of collisionless hydrodynamics are also obtained in the usual three-vector form.

The paper is organized as follows. In \Sec{sec:slicing}, we restate the $3+1$ formalism by amending the approach that was adopted in \Ref{ref:thorne82}. In \Sec{sec:motion}, we derive the equations of individual charged particles interacting with an electromagnetic field in space with an arbitrary metric. In \Sec{sec:vlasov}, we obtain the Vlasov equation in a number of equivalent forms. In \Sec{sec:hydro}, we derive the equations of collisionless hydrodynamics. In \Sec{sec:conclusions}, we summarize our main results. Supplementary calculations showing how our formalism relates to that in \Ref{ref:thorne82} are given in Appendix.

\section{Spacetime geometry} 
\label{sec:slicing}

In this section, we restate the $3+1$ formalism, which is based on the so-called slicing approach summarized in \Ref{ref:thorne82} (not to be confused with the $1+3$  formalism, which is based on the so-called threading approach \cite{arX:gourgoulhon07, ref:elst97}). Our purpose is to restate the known theory in a systematic, self-contained, and yet concise and simple form, relying on the reader's background in linear algebra rather than differential geometry; hence, the rules of index manipulation, still required, are included here. For reviews and other formulations, see \Refs{ref:thorne82, ref:jantzen92, ref:elst97, ref:bini97, ref:bini01, tex:bini98a, tex:bini98b, ref:baumgarte03, ref:sakai03, ref:katanaev06, ref:zanna07, arX:gourgoulhon07}.

\subsection{Metric tensor and index manipulation}
\label{sec:vecnot}

First, let us provide a brief introduction to the part of tensor analysis that will allow a reader, assumed familiar with linear algebra, to understand the rest of the paper (excluding the Appendix, which contains no new results). Suppose that the spacetime is equipped with a metric tensor $\foper{g}$ (the caret is to denote rank-two tensors), which then defines a scalar product for any given four-vectors $\fvec{X}$ and~$\fvec{Y}$,
\begin{gather}
\fvec{X} \cdot \fvec{Y} \equiv \foper{g}(\fvec{X}, \fvec{Y}),
\end{gather}
as a symmetric bilinear form. Consider a set of basis vectors $\fvec{e}_\mu$, where the Greek indexes span from $0$ to $3$. Then, each vector $\fvec{X}$ can be decomposed as $\fvec{X} = \fvec{e}_\mu X^\mu$ (and similarly for $\fvec{Y}$; summation over repeated indexes is assumed), yielding 
\begin{gather}\label{eq:scpr}
\fvec{X} \cdot \fvec{Y} = g_{\mu\nu} X^\mu Y^\nu,
\quad 
g_{\mu\nu} = g_{\nu\mu} \equiv \foper{g}(\fvec{e}_\nu, \fvec{e}_\mu).
\end{gather}
Unlike in the Euclidean (or Minkowski) space, the metric coefficients $g_{\mu\nu} = \fvec{e}_\mu \cdot \fvec{e}_\nu$ may not form a diagonal matrix, \ie $\fvec{e}_\mu$ are not orthogonal to each other. Hence, additional, ``dual'' basis vectors $\fvec{e}^\mu$ are introduced \cite{foot:dual}~via
\begin{gather}\label{eq:orth}
\fvec{e}^\mu \cdot \fvec{e}_\nu = \delta^\mu_\nu,
\end{gather}
which can also be understood as the definition of the ``mixed'' metric coefficients ${g^\mu}_\nu = {g_\nu}^\mu = \delta^\mu_\nu$. Then, the vector components $X^\mu$ in the original basis, or the so-called contravariant components, can be found as $X^\mu \equiv \fvec{e}^\mu \cdot \fvec{X}$. On the other hand, one may as well define the so-called covariant components as the vector components $X_\mu$ in the dual basis, as $X_\mu \equiv \fvec{e}_\mu \cdot \fvec{X}$; then,
\begin{gather}\label{eq:covcontra}
X_\mu = \fvec{e}_\mu \cdot \fvec{e}_\nu X^\nu = g_{\mu\nu} X^\nu.
\end{gather}
Finally, let us define the inverse metric tensor $\foper{g}^{-1}$ with the metric coefficients denoted as $g^{\mu\nu} = g^{\nu\mu}$ (the symmetry being inherited from that of $g_{\mu\nu}$); hence, from \Eq{eq:covcontra},
\begin{gather}
X^\mu = g^{\mu\nu} X_\nu.
\end{gather}
Then, \Eq{eq:scpr} can be equivalently put as
\begin{gather}
\fvec{X} \cdot \fvec{Y} = g_{\mu\nu} X^\mu Y^\nu = g^{\mu\nu} X_\mu Y_\nu = X^\mu Y_\mu = X_\mu Y^\mu.
\end{gather}

In addition to the four-vector space, we will also be dealing with a three-vector space (\Sec{sec:spmetric}), with objects to be denoted with bold ($\vec{X}$ instead of $\fvec{X}$). For those, the same rules of index manipulation apply, except with the metric tensor $\eta_{ij}$ (instead of $g_{\mu\nu}$) and Latin indexes (instead of Greek indexes), spanning from 1 to 3. For further reading on tensor analysis in application to the general relativity, one is referred, \eg to \Refs{book:misner77, book:weinberg, book:landau2}.

\subsection{Spacetime basis}
\label{sec:basis}

Suppose that time $t \equiv x^0$ (assuming the speed of light is equal to one) is defined as some function of spacetime location, such that constant-$t$ hypersurfaces $\Sigma_t$ are space-like. Introduce three arbitrary basis vectors $\fvec{e}_i(x^\mu)$ as tangent to these surfaces; hence the generalized coordinates $x^i$ in $\Sigma_t$, which we denote as space. Then, an arbitrary infinitesimal four-vector $d\fvec{x}$ is decomposed as 
\begin{gather}\label{eq:gendx0}
d\fvec{x} =  \fvec{e}_i\, dx^i + \fvec{e}_0 \, dt,
\end{gather}
where $\fvec{e}_0$ is the basis vector along the time axis, yet to be defined. Since the four-gradient $\del t$ is normal to $\Sigma_t$ (and thus orthogonal to $\fvec{e}_i$), \Eq{eq:gendx0} yields
\begin{gather}\label{eq:aux11}
\del t \cdot d\fvec{x} =   (\del t \cdot \fvec{e}_0)\,dt.
\end{gather}
On the other hand, 
\begin{gather}\label{eq:dt0}
dt = \frac{\pd t}{\pd x^\mu}\,dx^\mu = \del t \cdot d\fvec{x}.
\end{gather}
Then, from \Eq{eq:aux11}, one obtains
\begin{gather}\label{eq:delte0}
\del t \cdot \fvec{e}_0 = 1,
\end{gather}
meaning that $\del t$ is dual to $\fvec{e}_0$, \ie $ \del t = \fvec{e}^0$. 

Consider a normalized vector
\begin{gather}\label{eq:ngrad}
\fvec{n} = - \alpha \del t,
\end{gather}
where $\alpha$ is a scalar function such that
\begin{gather}\label{eq:ngrad2}
\fvec{n} \cdot \fvec{n} = - 1,
\end{gather}
and the sign is chosen assuming the metric signature
\begin{gather}\label{eq:signature}
(-,+,+,+)
\end{gather}
(That is, $\fvec{n}$ is the time-like unit normal to $\Sigma_t$.) Then, \Eq{eq:delte0} finally rewrites as 
\begin{gather}\label{eq:delte1}
\fvec{n} \cdot \fvec{e}_0 = -\alpha.
\end{gather}

\begin{figure*}
\centering
\includegraphics[width=0.75 \textwidth]{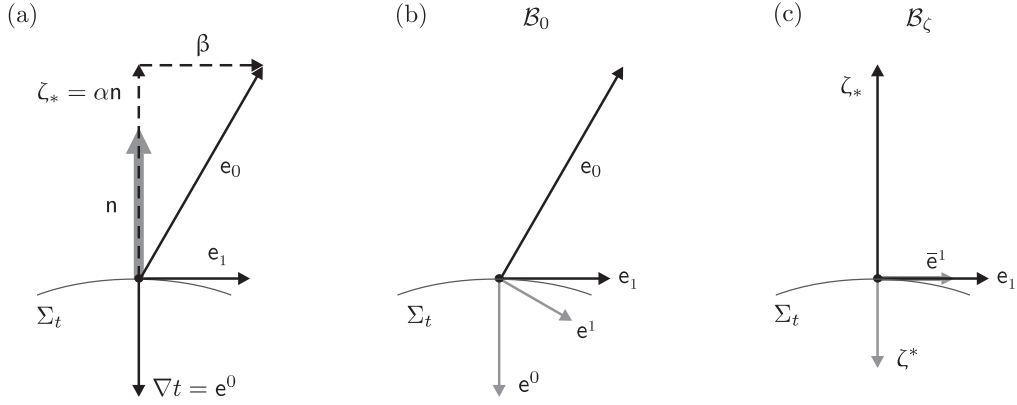}
\caption{Schematic of the spacetime bases (not in scale). (a)~Here $\fvec{n}$ is the unit normal to the space hypersurface $\Sigma_t$; $\fvec{e}_0$ is the basis vector that determines the time axis ($t$-axis); $\fvec{e}_1$ is the spatial basis vector (the other two spatial dimensions are not shown); $\upzeta_* = \alpha\fvec{n}$ determines $\zeta$-axis; $\alpha$ is the lapse function; $\upbeta$ is the shift vector. (b)~The vectors $(\fvec{e}_0,\fvec{e}_i)$ form the basis $\mc{B}_0$; also shown is the dual basis $(\fvec{e}^0,\fvec{e}^i)$. (c)~The vectors $(\upzeta_*, \fvec{e}_i)$ form the basis $\mc{B}_\zeta$; also shown is the dual basis $(\upzeta^*,\fvecp{e}^i)$.}
\label{fig:basis}
\end{figure*}

Equation \eq{eq:delte1} is the \textit{only} requirement on how $\fvec{e}_0$ should be defined. Although the time axis could be normal to $\Sigma_t$ (like $\fvec{n}$ and $\del t$), in general, $\fvec{e}_0$ can also have a component $\upbeta$ tangent to $\Sigma_t$. Thus, from \Eqs{eq:ngrad2} and \eq{eq:delte1}, the general form of this basis vector is [\Fig{fig:basis}(a)]
\begin{gather}\label{eq:e0gen}
\fvec{e}_0 = \alpha \fvec{n} + \upbeta,
\end{gather}
and the latter can be understood in two different ways. If $\fvec{e}_0$ is imposed, then functions $\alpha$ and $\upbeta$ are used to \textit{parameterize} the given basis $\mc{B}_0 \equiv (\fvec{e}_0,\fvec{e}_i)$. Alternatively, one may be allowed to choose $\fvec{e}_0$ as needed; in that case, $\alpha$ and $\upbeta$ are \textit{free} parameters, and it may be convenient to pick them differently depending on a problem of interest. 

To understand the physical meaning of the functions $\alpha$ and $\upbeta$, consider the coordinate form of~$\fvec{n}$ in $\mc{B}_0$:
\begin{align}
n^0 & \equiv \fvec{e}^0 \cdot \fvec{n} = \del t \cdot \fvec{n} = -(\fvec{n} \cdot \fvec{n})/\alpha = 1/\alpha, \\
n^i & \equiv \fvec{e}^i \cdot \fvec{n} = [\fvec{e}^i \cdot (\fvec{e}_0 - \upbeta)]/\alpha = - \beta^i/\alpha, \\
n_0 & \equiv \fvec{e}_0 \cdot \fvec{n} = (\alpha \fvec{n} + \upbeta) \cdot \fvec{n} = - \alpha, \\
n_i & \equiv \fvec{e}_i \cdot \fvec{n} = -\alpha\, (\fvec{e}_i \cdot \fvec{e}^0) = 0, 
\end{align}
or, in a compressed form,
\begin{gather}
n^\mu = (1/\alpha, -\beta^i/\alpha), \quad n_\mu = (-\alpha, 0_i)\label{eq:ncoord}
\end{gather}
[cf. \Fig{fig:basis}(a) and (b)]. Then, due to \Eq{eq:ngrad2}, one can treat $\fvec{n}(x^\mu)$ as the four-velocity of some observer at $x^\mu$, which we call, after \Ref{ref:thorne82}, a fiducial observer (FO). Introducing the FO proper time $\taufo$, one gets [\Eq{eq:ncoord}]
\begin{gather}
\alpha = d\taufo/dt.
\end{gather}
Therefore, $\alpha$ is called the ``lapse function''. Similarly, $\beta^i$ can be understood as minus the spatial velocity of FO, meaning that $\beta^i$ determine the rate at which the coordinate mesh on $\Sigma_t$ is shifting with respect to FO; thus, $\upbeta$ is called the ``shift vector''. Since the latter is not a physical velocity, it can be arbitrary, including superluminal, and therefore $\fvec{e}_0$ is not necessarily a time-like vector.

\subsection{Spatial metric}
\label{sec:spmetric}

To switch from a spacetime, where an arbitrary vector $\fvec{X}$ is decomposed as [see \Eq{eq:e0gen}]
\begin{gather}\label{eq:fvecdx}
\fvec{X} =  \fvec{e}_i X^i + (\alpha \fvec{n} + \upbeta)\, X^0,
\end{gather}
to the three-vector representation, construct the spatial vector space as follows. Introduce a symmetric tensor
\begin{gather}\label{eq:htensor}
h^\mu_\nu = \delta^\mu_\nu + n^\mu n_\nu,
\end{gather}
or $\foper{h} = \foper{I} + \fvec{n}\fvec{n}$ (where $\foper{I}$ is the unit tensor), to project $\fvec{X}$ on the plane tangent to~$\Sigma_t$:
\begin{gather}\label{eq:aux10}
\fvecp{X} \equiv \foper{h}\cdot \fvec{X} = \fvec{X} + (\fvec{n} \cdot \fvec{X})\,\fvec{n}.
\end{gather}
Since $\foper{h} \cdot \fvec{n} = 0$, directly from \Eq{eq:fvecdx} one can see that
\begin{gather}\label{eq:mc}
\proj{X}^\mu = (0, X^i + \beta^i X^0).
\end{gather}
For $\fvecp{X}$ is a tensor contraction, $\proj{X}^\mu$ transform as vector components by definition, and thus so do $\proj{X}^i$ (but not necessarily $X^i$). On the other hand, $\proj{X}^0 \equiv 0$; hence,
\begin{gather}
\vec{X} \equiv (\proj{X}^1,\proj{X}^2,\proj{X}^3)
\end{gather}
can be considered as a spatial three-vector.

Consider a length element in $\Sigma_t$:
\begin{gather}\label{eq:dxdx1}
d\vec{x} \cdot d\vec{x} \equiv d\proj{x}^i d\proj{x}_i = d\proj{x}^\mu d\proj{x}_\mu,
\end{gather}
where we used that $d\proj{x}^0 = 0$. Notice further that
\begin{gather}
d\fvecp{x} = \foper{h} \cdot d\fvecp{x},
\end{gather}
and thus $d\proj{x}_\mu = h_{\mu\nu} d\proj{x}^\nu$. Then, \Eq{eq:dxdx1} yields
\begin{gather}\label{eq:dxdx2}
d\vec{x} \cdot d\vec{x} = h_{\mu\nu} d\proj{x}^\mu d\proj{x}^\nu = \eta_{ij} d\proj{x}^i d\proj{x}^j,
\end{gather}
where we introduced $\eta_{ij} \equiv h_{ij}$ to distinguish the four-tensor $h_{\mu\nu}$ from its spatial part, which represents a symmetric three-tensor. From \Eq{eq:dxdx2}, it is then convenient to address $\eta_{ij}$ as the spatial metric. One hence \textit{defines} the mixed and contravariant metric tensors through
\begin{gather}\label{eq:invtdef}
\eta^i_j = \eta^{ik} \eta_{kj} = \delta^i_j.
\end{gather}
This allows raising and lowering indexes as
\begin{gather}\label{eq:3dindex}
\eta^{ij} \proj{X}_j = \proj{X}^i, \quad
\eta_{ij} \proj{X}^j = \proj{X}_i,
\end{gather}
where the covariant components satisfy
\begin{gather}\label{eq:covcomp}
\proj{X}_i = X_i,
\end{gather}
as flows from \Eq{eq:aux10} and $n_i = 0$. Hence, a three-vector scalar product can be defined as
\begin{gather}
\vec{X} \cdot \vec{Y} \equiv \eta_{ij} \proj{X}^i \proj{Y}^j = \eta^{ij} \proj{X}_i \proj{Y}_j = \proj{X}^i \proj{Y}_i = \proj{X}_i \proj{Y}^i.
\end{gather}

\subsection{Spacetime metric}
\label{sec:stmetric}

The three-vector components, \Eq{eq:mc}, can be understood as components of $\fvec{X}$ in the basis $\mc{B}_\zeta \equiv (\upzeta_*, \fvec{e}_i)$, with the dual basis being $(\upzeta^*, \fvecp{e}^i)$, where 
\begin{gather}
\upzeta_* = \alpha \fvec{n}, 
\quad 
\upzeta^* = - \alpha^{-1}\fvec{n},
\end{gather}
and $\fvecp{e}^i = \foper{h}\cdot \fvec{e}^i$ [\Fig{fig:basis}(c)]. Denoting the vector components in $\mc{B}_\zeta$ with underbars, one obtains, in particular, that
\begin{gather}\label{eq:uprojn}
\uproj{n}^\mu = (\alpha^{-1}, 0^i), \quad \uproj{n}_\mu = (-\alpha, 0_i),
\end{gather}
and, for vectors $\fvecp{X}$ projected on the plane tangent to $\Sigma_t$ [\Eq{eq:aux10}],
\begin{gather}\label{eq:uproj}
\uproj{\proj{X}}^\mu = (0, \proj{X}^i), \quad \uproj{\proj{X}}_\mu = (0, \proj{X}_i).
\end{gather}
In other words, the components of \textit{spatial} four-vectors are the same in $\mc{B}_0$ and $\mc{B}_\zeta$, and will not be distinguished from now on. Finally, from $d\fvec{x} =  \alpha \fvec{n}\, dt + d\fvecp{x}$ [see \Eq{eq:fvecdx}], we get
\begin{gather}\label{eq:fvecxn2}
d\fvec{x} =  \upzeta_*\,dt  + d\fvecp{x},
\end{gather}
so $d \zeta \equiv \upzeta^* \cdot d\fvec{x}$ equals $dt$. (Yet, when taking partial derivatives, one must distinguish $\zeta$-axis, along which $\proj{x}^i$ are fixed, from $x^0$-axis, or $t$-axis, along which $x^i$ are fixed.) Hence, an arbitrary spacetime interval can be put as
\begin{gather}\label{eq:dlfin}
d\fvec{x} \cdot d\fvec{x} = - \alpha^2 dt^2 + d\vec{x} \cdot d\vec{x}.
\end{gather}
Therefore, in the basis $\mc{B}_\zeta$, the metric tensor is represented by a block-diagonal matrix
\begin{gather}
g_{\mu\nu}
= \left(
\begin{array}{c @{\quad} c}
-\alpha^2 & 0\\
0 & \eta_{ij}
\end{array}
\right).\label{eq:gmetric}
\end{gather}

\section{Single particle motion}
\label{sec:motion}

In this section, we derive the motion equations for individual charged particles interacting with an electromagnetic field in space with an arbitrary metric $\eta_{ij}$. We consider both the coordinate and the vector form of these equations. Particularly, this explains how Eq.~(3.8) in \Ref{ref:thorne82}, similar to those, can be obtained without using the concept of a covariant derivative (also see Appendix). Additional information on the Hamiltonian formalism for the particle motion in curved spacetime can be found in \Ref{ref:cognola86}.

\subsection{Canonical equations}
\label{sec:canonic}

Consider a particle with mass $m$ and charge $q$ interacting with an electromagnetic four-potential $\fvec{A}$, so the particle action $S$ reads as (see, \eg \Refs{ref:dewitt66, ref:cognola86})
\begin{gather}\label{eq:action}
S = \int\left( -m\,\sqrt{-d\fvec{x} \cdot d\fvec{x}} + q\kpt{1} \fvec{A} \cdot d\fvec{x} \right).
\end{gather}
Introduce the particle Lagrangian $L$ through $S = \int L\,dt$. Hence, from \Eq{eq:dlfin}, one gets
\begin{gather}\label{eq:bl}
L = - m \varkappa^{-1} + q A_j (\proj{v}^j - \beta^j) + q A_0,
\end{gather}
where the velocity components read as [\Eq{eq:mc}]
\begin{gather}\label{eq:twovel}
\proj{v}^i \equiv d\proj{x}^i/dt = v^i + \beta^i,
\end{gather}
and, with $\vec{v} \cdot \vec{v} \equiv \vec{v}^2$, $\varkappa$ is given by
\begin{gather}\label{eq:varkap}
\varkappa^{-1} = \sqrt{\alpha^2 - \vec{v}^2}.
\end{gather}
(As before, the bold symbol $\vec{v}$ denotes a three-vector with components $\proj{v}^i$, not $v^i \equiv dx^i/dt$.) Then, one puts \Eq{eq:bl} in the following form:
\begin{gather}
L = - m \sqrt{\alpha^2 - \vec{v}^2} + q \vec{v} \cdot \vec{A} + qA_\zeta,
\end{gather}
where $A_j = \proj{A}_j$ [\Eq{eq:covcomp}] and 
\begin{gather}
A_\zeta \equiv \upzeta_* \cdot \fvec{A} = A_0 - \beta^j A_j
\end{gather}
is the covariant component of $\fvec{A}$ along $\zeta$-axis. 

The three canonical momenta $\proj{P}_i$, completing the canonical pairs $(\proj{x}^i, \proj{P}_i)$, are defined as $\proj{P}_i = \pd L/\pd \proj{v}^i$,~or
\begin{gather}\label{eq:cpi0}
\proj{P}_i = m \varkappa \eta_{ij} \proj{v}^j + q\proj{A}_i.
\end{gather}
Using \Eqs{eq:3dindex} and \eq{eq:covcomp}, one obtains then:
\begin{gather}
\proj{P}_i = m \varkappa \proj{v}_i + q\proj{A}_i = m \varkappa v_i + qA_i.
\end{gather}
Thus, the three $\proj{P}_i$ given by \Eq{eq:cpi0} are equal to the covariant components of the corresponding four-vector~$\fvec{P}$. Similarly, the kinetic momenta,
\begin{gather}\label{eq:kinmom}
\proj{p}_i = \proj{P}_i - q\proj{A}_i,
\end{gather}
equal
\begin{gather}\label{eq:cpt}
\proj{p}_i = m \varkappa \eta_{ij}\proj{v}^j = m \varkappa v_i, 
\end{gather}
and thus coincide with the corresponding components of the particle four-momentum $\fvec{p}$. Therefore, the formally introduced variables $\proj{P}_i$ and $\proj{p}_i$ satisfy (unlike $P_i$ and $p_i$) the index manipulation rules introduced in \Sec{sec:spmetric}.

Hence, invert \Eq{eq:cpt} as
\begin{gather}
\proj{v}^i = \frac{\eta^{ij}\proj{p}_j}{m\varkappa}.
\end{gather}
This allows the particle Hamiltonian,
\begin{gather}\label{eq:3dham}
H = \proj{P}_i \proj{v}^i - L =  \alpha^2 m \varkappa - q  A_\zeta,
\end{gather}
to be expressed as a function of $\proj{p}_i$. Specifically,
\begin{gather}\label{eq:gammaf}
\varkappa = \gamma/\alpha, \quad \gamma =\sqrt{1 + \vec{p}^2/m^2},
\end{gather}
thus, $H$ rewrites as
\begin{gather}
H = \alpha m \gamma  - q  A_\zeta,  \label{eq:3dham2}
\end{gather}
where, in terms of the canonical momentum $\vec{P}$, one has
\begin{gather}
\gamma = \sqrt{1 + (\vec{P} - q\vec{A})^2/m^2}.
\end{gather}
Finally note that, since one can also put the latter as
\begin{gather}\label{eq:gammaV}
\gamma = \frac{1}{\sqrt{1-\vec{V}^2}},
\end{gather}
where $\vec{V}$ is the particle velocity as seen by FO,
\begin{gather}\label{eq:V}
\vec{V} \equiv \frac{\vec{v}}{\alpha} = \frac{1}{\alpha}\,\frac{d\vec{x}}{dt} = \frac{d\vec{x}}{d\taufo},
\end{gather}
one can understand $\gamma$ as the particle Lorentz factor as measured by FO.

The canonical equations are now obtained as follows. For the coordinates, $d\proj{x}^i/dt = \pd H/\pd \proj{P}_i$ yield the already known equation
\begin{gather}
\frac{d\proj{x}^i}{dt} = \frac{\alpha \proj{p}^i}{\gamma m} = \proj{v}^i.\label{eq:vvsp}
\end{gather}
For the canonical momenta, one has $d\proj{P}_i/dt = - \pd H/\pd \proj{x}^i$, or
\begin{gather}
\frac{1}{\alpha}\,\frac{d\proj{P}_i}{dt} 
= m \gamma \grav_i 
+ \frac{\pd \eta_{j k}}{\pd \proj{x}^i}\frac{\proj{p}^j \proj{p}^k}{2m \gamma} 
+ \frac{q}{\alpha}\left(
\frac{\pd \proj{A}_\zeta}{\pd \proj{x}^i} + \frac{\pd \proj{A}_j}{\pd \proj{x}^i}\,\proj{v}^j
\right),
\end{gather}
where we substituted
\begin{gather}
\frac{\pd \eta^{jm}}{\pd \proj{x}^i} = - \eta^{\ell m}\eta^{jk}\,\frac{\pd \eta_{k \ell}}{\pd \proj{x}^i}
\end{gather}
[from differentiating \Eq{eq:invtdef}] and introduced $\grav_i$ (not to be confused with the metric tensor $\foper{g}$) as components of the three-vector
\begin{gather}
\vecgrav = - \vec{\del} \ln \alpha.
\end{gather}
(Unlike $\del$, which is the gradient in the 4D space, the bold symbol $\vec{\del}$ denotes the gradient in the 3D space.) Using \Eq{eq:kinmom}, one then gets for the kinetic momenta:
\begin{gather}\label{eq:grav0}
\frac{1}{\alpha}\,\frac{d\proj{p}_i}{dt} 
= m \gamma \grav_i 
+ \frac{\pd \eta_{j k}}{\pd \proj{x}^i}\frac{\proj{p}^j \proj{p}^k}{2m \gamma} + \proj{\Lambda}_i,
\end{gather}
with $\proj{\Lambda}_i$ being the Lorentz force:
\begin{gather}\label{eq:lambda}
\proj{\Lambda}_i = \frac{q}{\alpha}\left[
\frac{\pd \proj{A}_\zeta}{\pd \proj{x}^i} - \frac{\pd \proj{A}_i}{\pd \zeta}+ \left(\frac{\pd \proj{A}_j}{\pd \proj{x}^i} - \frac{\pd \proj{A}_i}{\pd \proj{x}^j}\right)\proj{v}^j
\right].
\end{gather}

\subsection{Lorentz force}
\label{sec:lorentz}

Consider expressing $\proj{\Lambda}_i$ in terms of
\begin{gather}\label{eq:emf1}
F_{\mu\nu} = \frac{\pd A_{\nu}}{\pd x^\mu} - \frac{\pd A_{\mu}}{\pd x^\nu},
\end{gather}
which falls under the definition of a tensor (see, \eg Sec.~83 in \Ref{book:landau2} or Sec.~4.2 in \Ref{book:weinberg}). By analogy with the Minkowski spacetime, write this so-called electromagnetic tensor (Sec.~90 in \Ref{book:landau2}) as
\begin{gather}\label{eq:emf2}
F_{\mu\nu} = n_\mu \proj{E}_\nu - n_\nu\proj{E}_\mu  + \epsilon_{\lambda\mu\nu\kappa} \proj{B}^\kappa n^\lambda.
\end{gather}
Here $\proj{\fvec{E}}$ and $\proj{\fvec{B}}$ are four-vectors with zero time components (hence the bars), $\epsilon_{\lambda\mu\nu\kappa}$ is the permutation pseudotensor:
\begin{align}
\epsilon_{\lambda\mu\nu\kappa} =  \sqrt{-g}\,[\lambda\mu\nu\kappa],
\end{align}
$[\lambda\mu\nu\kappa]$ is the permutation symbol, and
\begin{align} \label{eq:det}
g \equiv \det g_{\mu\nu} = - \alpha^2 \eta, \quad \eta \equiv \det \eta_{ij},
\end{align}
the equality flowing from \Eq{eq:gmetric}. [This form ensures that $F_{\mu\nu}$ is indeed a tensor and introduces the exact amount of free parameters (the six nonzero components of $\proj{\fvec{E}}$ and $\proj{\fvec{B}}$) to define an antisymmetric matrix like $F_{\mu\nu}$.]

In terms of $F_{\mu\nu}$, \Eq{eq:lambda} rewrites as
\begin{gather}\label{eq:lorentz2}
\proj{\Lambda}_i = \frac{q}{\alpha}\left(\uproj{F}_{i\zeta} + \uproj{F}_{ij}\proj{v}^j\right),
\end{gather}
where $\uproj{F}_{\mu\nu}$ are the corresponding components of $F_{\mu\nu}$ in the basis $\mc{B}_\zeta$. From \Eq{eq:emf2}, one gets
\begin{gather}
\uproj{F}_{i\zeta} = - \uproj{n}_\zeta \proj{E}_i  + \epsilon_{\lambda i \zeta j} \proj{B}^j \uproj{n}^\lambda,
\end{gather}
where we utilized \Eq{eq:uproj} for the spatial four-vectors. Employing \Eq{eq:uprojn} in the form
\begin{gather}
\uproj{n}_i = 0, \quad \uproj{n}_\zeta = -\alpha, \quad \uproj{n}^\lambda = \alpha^{-1} \delta^\lambda_\zeta,
\end{gather}
one gets
\begin{gather}
\uproj{F}_{i\zeta} = \alpha \proj{E}_i + \alpha^{-1}\epsilon_{\zeta i \zeta j} \proj{B}^j = \alpha \proj{E}_i,
\end{gather}
because $[\zeta i \zeta j] = 0$. Similarly,
\begin{multline}
\uproj{F}_{ij} 
 = \epsilon_{\lambda i j k} \proj{B}^k \uproj{n}^\lambda 
 = \alpha^{-1}\sqrt{-g}\,[\zeta ijk] \proj{B}^k \\
 = \sqrt{\eta}\,[\zeta ijk] \proj{B}^k
 = \sqrt{\eta}\,[ijk] \proj{B}^k
 = \epsilon_{ijk} \proj{B}^k.
\end{multline}
Here we use $\epsilon_{ijk} = \sqrt{\eta}\,[ijk]$ to define the permutation three-pseudotensor $\epsilon_{ijk}$. Unlike Eq.~(16) in \Ref{ref:baumgarte03}, this is a standard definition of $\epsilon_{ijk}$, which automatically ensures that the Lorentz force has a three-vector form.

Finally, \Eq{eq:lorentz2} rewrites~as
\begin{gather}\label{eq:lorentz3}
\proj{\Lambda}_i = q\left(\proj{E}_i + \alpha^{-1}\epsilon_{ijk} \proj{v}^j \proj{B}^k\right).
\end{gather}
With the definition \eq{eq:V}, one hence obtains that $\proj{\Lambda}_i$ can be regarded as covariant components of the three-vector
\begin{gather}\label{eq:lorentz4}
\vec{\Lambda} = q\left(\vec{E} + \vec{V} \times \vec{B}\right).
\end{gather}
Equation \eq{eq:lorentz4} is similar to that in the Minkowski space. However, notice the difference between $\vec{V}$ that enters here and the velocity $\vec{v} = d\vec{x}/dt$ [\Eq{eq:vvsp}]; in particular, see Appendix for comparison with \Ref{ref:holcomb89}. Notice also that, within the three-vector formalism that we adopt, the shift vector $\vec{\beta}$ does not explicitly enter the above derivation of the Lorentz force, unlike in \Ref{ref:baumgarte03}. It can be reintroduced, though, by substituting $\proj{v}^j = v^j + \beta^j$ [\Eq{eq:twovel}]. Contrary to \Ref{ref:zanna07}, the so-called transport velocity $v^j \equiv dx^j/dt$ here is also a three-vector, \eg because it equals the difference of three-vectors $\proj{v}^j$ and $\beta^j$.

\subsection{Metric-caused forces}
\label{sec:grav}

Now let us revert to \Eq{eq:grav0} and calculate the effect due to the metric $\eta_{ij}$. First, raise the index using
\begin{gather}
\frac{d\proj{p}^\ell}{dt} 
 = \frac{d(\eta^{\ell i} \proj{p}_i)}{dt} 
 = \eta^{\ell i}\frac{d\proj{p}_i}{dt} 
 + \left( \frac{\pd \eta^{\ell i}}{\pd \zeta} + \proj{v}^k \frac{\pd \eta^{\ell i}}{\pd \proj{x}^k} \right) \proj{p}_i.
\end{gather}
Substituting \Eq{eq:vvsp} and employing \Eq{eq:invtdef}, one obtains (similarly to Sec.~87 in \Ref{book:landau2}):
\begin{gather}
\frac{1}{\alpha}\,\frac{d\proj{p}^\ell}{dt} 
 = - \frac{\proj{\Gamma}^\ell_{kj} \proj{p}^j\proj{p}^k}{m\gamma} + m \gamma \grav^\ell - \frac{\eta^{\ell i}}{\alpha}\,\frac{\pd \eta_{ij}}{\pd \zeta}\,\proj{p}^j + \proj{\Lambda}^\ell,
\end{gather}
where $\proj{\Gamma}^\ell_{kj} = \proj{\Gamma}^\ell_{jk}$ are given by
\begin{gather}\label{eq:gammah}
\proj{\Gamma}^\ell_{jk} = \frac{\eta^{\ell i}}{2}\left(\frac{\pd \eta_{ik}}{\pd \proj{x}^j} +
\frac{\pd \eta_{ij}}{\pd \proj{x}^k}  - \frac{\pd \eta_{jk}}{\pd \proj{x}^i}\right),
\end{gather}
also known as connection coefficients, or Christoffel symbols, associated with the metric $\eta_{ij}$ (Chap.~4 in \Ref{book:weinberg}). Introduce what is called the extrinsic curvature of the spatial surfaces $\Sigma_t$ as \cite{foot:liecurv}
\begin{gather}\label{eq:kij}
\proj{K}_{ij} = - \frac{1}{2\alpha}\,\frac{\pd \eta_{ij}}{\pd \zeta}
\end{gather}
(see, \eg \Ref{arX:gourgoulhon07} or Sec.~21.5 in \Ref{book:misner77}). Then,
\begin{gather}\label{eq:sceqpi}
\frac{1}{\alpha}\,\frac{d\proj{p}^\ell}{dt} 
 =  - \frac{\proj{\Gamma}^\ell_{kj} \proj{p}^j\proj{p}^k}{m\gamma} + m \gamma \grav^\ell + 2 \proj{K}^\ell_j \proj{p}^j + \proj{\Lambda}^\ell,
\end{gather}
where $\proj{K}^\ell_j = \eta^{\ell i} \proj{K}_{ij}$, in agreement with the standard rules of spatial index manipulation [\Eqs{eq:3dindex}].

In \Eq{eq:sceqpi}, the first term on the right-hand side is due to the generally non-Euclidean form of $\eta_{ij}$, the second one is the gravity force, and the third one is due to the curvature of $\Sigma_t$ considered as a \textit{subspace} of spacetime (while $\eta_{ij}$, being the \textit{own} metric of $\Sigma_t$, may or may not exhibit an intrinsic curvature). The vector equation that we will now derive illustrates these forces in further~detail.

\subsection{Vector equation}
\label{sec:vectormotion}

It can be shown (Sec.~8.5 in \Ref{book:misner77}) that
\begin{gather}\label{eq:gammahe}
\proj{\Gamma}^\ell_{kj} = \vec{e}^\ell \cdot \frac{\pd\kpt{.5} \vec{e}_k}{\pd \proj{x}^j},
\end{gather}
where $\vec{e}_k$ are the basis three-vectors tangent to $\Sigma_t$, and $\vec{e}^\ell$ are those of the dual three-vector basis. Then, 
\begin{multline}
\frac{1}{\alpha}\,\frac{d\proj{p}^\ell}{dt} + \frac{\proj{\Gamma}^\ell_{kj} \proj{p}^j\proj{p}^k}{m\gamma} 
 = \frac{\vec{e}^\ell}{\alpha} \cdot\left[\vec{e}_k\,\frac{d\proj{p}^k}{dt} + \proj{p}^k (\vec{v} \cdot \vec{\del}) \vec{e}_k \right] \\
 = \frac{1}{\alpha}\left[\vec{e}^\ell \cdot \frac{d(\vec{e}_k \proj{p}^k)}{dt} - \vec{e}^\ell \cdot \frac{\pd \vec{e}_k}{\pd \zeta}\,\proj{p}^k\right].
\end{multline}
The first term in the square brackets equals the $\ell$th projection of $d\vec{p}/dt$, whereas the second one can be represented as follows. First, notice that the expression
\begin{gather}
\vec{e}^\ell \cdot \frac{\pd \vec{e}_k}{\pd \zeta} = \fvecp{e}^\ell \cdot \frac{\pd \fvecp{e}_k}{\pd \zeta}
\end{gather}
coincides with the connection coefficient $\Gamma^\ell_{k\zeta}$ associated with the 4D metric $g_{\mu\nu}$ (rather than $\eta_{ij}$), written in the basis $\mc{B}_\zeta$. Similarly to \Eq{eq:gammah}, we can write then
\begin{gather}
\Gamma^\ell_{k\zeta} = \frac{\uproj{g}^{\ell \lambda}}{2}\left(
\frac{\pd \uproj{g}_{\lambda \zeta}}{\pd \uproj{x}^k} + \frac{\pd \uproj{g}_{\lambda k}}{\pd \uproj{x}^\zeta}  - \frac{\pd \uproj{g}_{k\zeta}}{\pd \uproj{x}^\lambda}\right),
\end{gather}
where the underbars show that the expression is evaluated in $\mc{B}_\zeta$. From \Eq{eq:gmetric}, both covariant and contravariant metric elements with mixed (\ie space-time) coefficients are zero. Then, since $\uproj{x}^\zeta \equiv \zeta$, one gets
\begin{multline}
\Gamma^\ell_{k\zeta} = 
 \frac{\uproj{g}^{\ell j}}{2}
  \left(\frac{\pd \uproj{g}_{j \zeta}}{\pd \uproj{x}^k} + \frac{\pd \uproj{g}_{j k}}{\pd \zeta}  - \frac{\pd \uproj{g}_{k\zeta}}{\pd \uproj{x}^j}\right) \\
 = \frac{1}{2}\,\uproj{g}^{\ell j}\frac{\pd \uproj{g}_{j k}}{\pd \zeta}
 = \frac{1}{2}\,\eta^{\ell j}\frac{\pd \eta_{j k}}{\pd \zeta}.
\end{multline}
Comparing this with \Eq{eq:kij}, we obtain
\begin{gather}\label{eq:eijK}
\vec{e}^\ell \cdot \frac{\pd \vec{e}_k}{\pd \zeta} = - \alpha \proj{K}^\ell_k.
\end{gather}
Hence, \Eq{eq:sceqpi} rewrites as the following vector equation, independent of the spatial basis:
\begin{gather}\label{eq:vec1}
\frac{1}{\alpha}\,\frac{d\vec{p}}{dt} 
 =  m \gamma \vecgrav + \oper{K} \cdot \vec{p} +  q\left(\vec{E} + \vec{V} \times \vec{B}\right).
\end{gather}
[Note the unit coefficient in the term $\oper{K} \cdot \vec{p}$, unlike in \Eq{eq:sceqpi}.] For an alternative representation of \Eq{eq:vec1} and comparison with similar representations found in literature, see Appendix.

\section{Vlasov equation}
\label{sec:vlasov}

There are several ways to introduce the distribution function $f$ describing the 3D motion in $\Sigma_t$ that naturally extrapolate the one from the Minkowski spacetime. In this section, we will show that those definitions are equivalent; yet, the equation for $f$ can take various forms. (Notice also that, when electromagnetic interactions are included, one may need to close the Vlasov theory with the Maxwell's equations. For those, see, \eg Eqs.~(3.4) in \Ref{ref:thorne82} and also our Appendix for the notation.)

\subsection{Liouville theorem}

First, let us define $f$ as the particle density in the 6D phase space, that is,
\begin{gather}\label{eq:fdef}
f = d \mc{N}/d \Omega,
\end{gather}
where $d \mc{N}$ is the number of particles in the phase volume element $d \Omega$. In this case, the Vlasov equation can be derived immediately from the Liouville theorem (Sec.~3 in \Ref{book:landau10}). Namely, the latter says that $d\Omega$ is conserved; then, since $d\mc{N}$ is also constant, one gets
\begin{gather}\label{eq:vlasovgen}
\frac{\mathit{df}}{dt} = 0,
\end{gather}
that is, $f$ is conserved along the particle 6D trajectories. 

Yet these trajectories do not have to be expressed through canonical variables. For example, one may consider $f$ as a function of $\proj{x}^i$ and $\proj{p}^i$. Then, in agreement with \Ref{ref:debbasch09b}, \Eq{eq:vlasovgen} rewrites like that for the Minkowski spacetime (see, \eg Chap.~8 of \Ref{book:stix}),~\ie
\begin{gather}\label{eq:vl1}
\frac{\pdf}{\pd \zeta} + \frac{d\proj{x}^i}{dt}\,\frac{\pdf}{\pd \proj{x}^i} + \frac{d\proj{p}^i}{dt}\,\frac{\pdf}{\pd \proj{p}^i} = 0,
\end{gather}
and one can use \Eqs{eq:vvsp} and \eq{eq:sceqpi} for $d\proj{x}^i/dt$ and $d\proj{p}^i/dt$ to close it.

\subsection{Divergence form}
\label{eq:divform}

To connect $f$ with \textit{measurable} quantities, $d \Omega$ in \Eq{eq:fdef} is derived as follows. From \Sec{sec:canonic}, it flows that the canonical variables are contravariant components $\proj{x}^i$ and the canonical momenta are covariant components $\proj{P}_i$. Since
\begin{gather}
\left|\frac{\pd (\proj{x}^i,\proj{P}_i)}{\pd (\proj{x}^j,\proj{p}_j)}\right| = 1,
\end{gather}
one can also use kinetic momenta, though; then,
\begin{gather}\label{eq:Pi}
d\Omega = d^3\proj{x}^* d^3\proj{p}_*,
\end{gather}
where $d^3\proj{x}^* \equiv d\proj{x}^1\, d\proj{x}^2\, d\proj{x}^3$ and $d^3\proj{p}_* \equiv d\proj{p}_1\, d\proj{p}_2\, d\proj{p}_3$ \cite{foot:dform}. (From now on, asterisk denotes whether upper or lower indexes are assumed.)

Hence, one needs to find how $d^3\proj{x}^*$ and $d^3\proj{p}_*$ are connected with the invariant physical volumes in the coordinate and momentum spaces \cite{ref:debbasch09a}. Since both of those are vector spaces with the metric $\eta_{ij}$, one can write
\begin{gather}
d \mc{V}_{\vec{x}} = \sqrt{\eta}\, d^3\proj{x}^*,
\quad
d \mc{V}_{\vec{p}} = \sqrt{\eta}\, d^3\proj{p}^*.
\end{gather}
Thus, 
\begin{gather}
d\Omega = \frac{1}{\eta}\,\left| 
\frac{\pd (\proj{x}^i,\proj{p}_i)}{\pd (\proj{x}^j,\proj{p}^j)} 
\right|\, d \mc{V}_{\vec{x}} d \mc{V}_{\vec{p}}.
\end{gather}
Since $\proj{p}_i = \eta_{ij}\proj{p}^j$, the Jacobian here equals $\eta$; therefore, 
\begin{gather}
d\Omega = d \mc{V}_{\vec{x}} d \mc{V}_{\vec{p}}.
\end{gather}
This means that $f$, originally defined as the phase-space density [\Eq{eq:fdef}], is also the density in $(\proj{x}^i, \proj{p}^i)$ space:
\begin{gather}\label{eq:fdef2}
f = \frac{d\mc{N}}{d \mc{V}_{\vec{x}} d \mc{V}_{\vec{p}}},
\end{gather}
in agreement with \Ref{ref:debbasch09a}.

This result allows yet another representation of the Vlasov equation, which is derived as follows. First, rewrite \Eq{eq:fdef2} as
\begin{gather}
\mathit{\eta f} = \frac{d\mc{N}}{d^3\proj{x}^*d^3\proj{p}^*},
\end{gather}
which means that $\mathit{\eta f}$ can be considered as the density $f_\eta$ in $(\proj{x}^i, \proj{p}^i)$ space, if the latter is assigned the Euclidean volume form (so the elementary volume is not $d \mc{V}_{\vec{x}} d \mc{V}_{\vec{p}}$ but rather $d^3\proj{x}^*\,d^3\proj{p}^*$). Then, the particle current along $\proj{x}^i$-axis is $j^i_{\vec{x}} = f_\eta d\proj{x}^i/dt$, and the current along $\proj{p}^i$-axis is $j^i_{\vec{p}} = f_\eta d\proj{p}^i/dt$, yielding that the particle conservation law reads as usual,
\begin{gather}
\frac{\pdf_\eta}{\pd \zeta} + 
\frac{\pd j^i_{\vec{x}}}{\pd \proj{x}^i} + 
\frac{\pd j^i_{\vec{p}}}{\pd \proj{p}^i} = 0
\end{gather}
(because the 6D space is considered Euclidean). Then, using that $f_\eta = \mathit{\eta f}$, one gets
\begin{gather}\label{eq:divform1}
\frac{\pd\left(\mathit{\eta f}\right)}{\pd \zeta} + 
\frac{\pd}{\pd \proj{x}^i}\left(\frac{d\proj{x}^i}{dt}\,\mathit{\eta f} \right) + 
\frac{\pd}{\pd \proj{p}^i}\left(\frac{d\proj{p}^i}{dt}\,\mathit{\eta f} \right) = 0,
\end{gather}
which we henceforth call the divergence form of the Vlasov equation. Equation \eq{eq:divform1}, considered in combination with \Eq{eq:vvsp} for $d\proj{x}^i/dt$ and \Eq{eq:sceqpi} for $d\proj{p}^i/dt$ [and \Eq{eq:lorentz3} for $\proj{\Lambda}^i$], represents the main result of this paper. In \Sec{sec:hydro}, it will also be used to yield three-vector equations of collisionless plasma hydrodynamics.

\subsection{Other representations}

The variables $(\proj{x}^i, \proj{p}^i)$ are natural for describing dynamics on spatial hypersurfaces $\Sigma_t$, because they allow for vector interpretation and are also self-contained (\eg $p^0$ and $p_0$ do not need to be considered). Yet, let us show how our formalism extrapolates to the original variables $(x^i, p^i)$, particularly, to compare with \Ref{ref:debbasch09a}. 

In the form \eq{eq:vlasovgen}, the Vlasov equation in variables $(x^i, p^i)$ can be written immediately~as 
\begin{gather}\label{eq:vl2}
\frac{\pdf}{\pd t} + \frac{d x^i}{dt}\,\frac{\pdf}{\pd x^i} + \frac{d p^i}{dt}\,\frac{\pdf}{\pd p^i} = 0,
\end{gather}
in agreement with \Ref{ref:debbasch09b}. Correspondingly, the equations for $dx^i/dt$ and $dp^i/dt$ can be obtained, \eg from \Eqs{eq:vvsp} and \eq{eq:sceqpi}, since
\begin{gather}
\frac{dx^i}{dt} = \frac{d\proj{x}^i}{dt} - \beta^i, 
\quad 
\frac{dp^i}{dt} = \frac{d\proj{p}^i}{dt} - \frac{d(\beta^i p^0)}{dt}
\end{gather}
[cf. \Eqs{eq:mc} and \eq{eq:twovel}]. Here $p^0$ is given by
\begin{gather}\label{eq:p0}
 p^0 = m\,\frac{dx^0}{d\tau} \equiv m\,\frac{dt}{d\tau} = m\varkappa = \frac{m\gamma}{\alpha},
\end{gather}
$\tau$ is the particle proper time, and [see \Eq{eq:gammaf}]
\begin{gather}\label{eq:anotherg}
\gamma = \sqrt{1 + \eta_{ij}\proj{p}^i\proj{p}^j /m^2} = \sqrt{1 + \eta^{ij}p_ip_j /m^2},
\end{gather}
the latter equality (to be used below) being due to
\begin{gather}\label{eq:pipi}
\proj{p}_i = p_i.
\end{gather}

Similarly, the divergence form is derived as follows. First, using \Eq{eq:pipi}, rewrite \Eq{eq:Pi} in the form
\begin{gather}\label{eq:pi2}
d\Omega 
= \left|\frac{\pd (\proj{x}^i, p_i)}{\pd (x^j,p_j)}\right| d^3x^* d^3 p_* 
= d^3x^* d^3 p_*
\equiv r\,d^3x^* d^3 p^*,
\end{gather}
where $r$ is the following Jacobian:
\begin{gather}\label{eq:r1}
\frac{1}{r} =\left| \frac{\pd p^j}{\pd p_i}\right|.
\end{gather}
Combine \Eq{eq:pi2} with \Eq{eq:fdef}, so one gets
\begin{gather}
\mathit{rf} = \frac{d\mc{N}}{d^3 x^*d^3 p^*}.
\end{gather}
Then, by analogy with \Eq{eq:divform1}, we immediately obtain
\begin{gather}\label{eq:divform2}
\frac{\pd\left(\mathit{rf}\right)}{\pd t} + 
\frac{\pd}{\pd x^i}\left(\frac{dx^i}{dt}\,\mathit{rf} \right) + 
\frac{\pd}{\pd p^i}\left(\frac{dp^i}{dt}\,\mathit{rf} \right) = 0,
\end{gather}
again in agreement with \Ref{ref:debbasch09a}.

Now let us show how the expression for $r$ is derived (without introducing the ``mass shell'' used in \Ref{ref:debbasch09a}). First, employ \Eq{eq:pipi}, yielding
\begin{gather}\label{eq:dpdpd1}
 \frac{\pd p^j}{\pd p_i} = \frac{\pd}{\pd \proj{p}_i}\left(\proj{p}^j - \beta^j p^0\right) = \eta^{ji} - \beta^j\,\frac{\pd p^0}{\pd p_i}.
\end{gather}
Then, using \Eq{eq:p0} together with \Eq{eq:anotherg}, one gets $\pd p^0/\pd p_i = \proj{v}^i/\alpha^2$, so \Eq{eq:dpdpd1} rewrites as
\begin{gather}\label{eq:dpdpd2}
 \frac{\pd p^j}{\pd p_i} = \eta^{ji} - \frac{\proj{v}^i\beta^j }{\alpha^2}  = \eta^{ji} {w^j}_i, \quad {w^j}_i = \delta^j_i - \frac{v_i\beta^j}{\alpha^2},
\end{gather}
and therefore $1/r = w/\eta$, where $w \equiv \det {w^j}_i$. One of the ways to find the determinant of ${w^j}_i$, which is a $3 \times 3$ matrix, is through a brute-force calculation. A somewhat more elegant (and independent of the number of dimensions, albeit longer) way would be to see that ${w^j}_i$ is a tensor of rank $(1,1)$, and thus $w$ is independent of the (spatial) basis. Hence, one can consider the basis such that $\vec{\beta}$ points, say, along $z$-axis. Then, $\beta^j = \delta^j_z \beta^z$, in which case one immediately finds $w = 1 - v_z\beta^z /\alpha^2$. In the invariant form, this result is expressed through the scalar product of $\vec{v}$ and $\vec{\beta}$; therefore,
\begin{gather}\label{eq:r2}
 r = \frac{\eta}{1 - \vec{v} \cdot {\vec{\beta}}/\alpha^2}.
\end{gather}
[When $\vec{\beta} = 0$, one gets $r = \eta$, and \Eq{eq:divform2} becomes equivalent to \Eq{eq:divform1}, because $(\proj{x}^i, \proj{p}^i)$ are then the same as $(x^i, p^i)$ and $t$-axis coincides with $\zeta$-axis.] Using that $p_0 = g_{0\mu}p^\mu$, with $g_{\mu\nu}$ expressed \cite{arX:gourgoulhon07} in the basis $\mc{B}_0$ [rather than $\mc{B}_\zeta$, as in \Eq{eq:gmetric}] and \Eqs{eq:p0} and \eq{eq:det}, one can also rewrite \Eq{eq:r2}~as
\begin{gather}
r =  \frac{p^0}{p_0}\,\det g_{\mu\nu}.
\end{gather}
Hence, our result agrees \cite{foot:sign} with that in \Ref{ref:debbasch09a}.

\section{Hydrodynamic equations}
\label{sec:hydro}

Finally, let us consider moments of the Vlasov equation to obtain equations of collisionless hydrodynamics. To preserve the vector form of the dynamic equations, we use $(\proj{x}^i, \proj{p}^i)$ variables. Correspondingly, the average of an arbitrary function $\chi$ over the momentum distribution is defined as
\begin{gather}
\favr{\chi} = \frac{1}{N} \int \chi f\,d\mc{V}_{\vec{p}} = \frac{\sqrt{\eta}}{N}  \int \chi f\,d^3\proj{p}^*,
\end{gather}
where $N$ stands for the particle density in the 3D space:
\begin{gather}
N = \int f\,d\mc{V}_{\vec{p}} = \sqrt{\eta}\int f\,d^3\proj{p}^*
\end{gather}
(not to be confused with the particle \textit{proper} density \cite{arX:gourgoulhon07}).

Integrating \Eq{eq:divform1} over $d^3\proj{p}^*$ yields
\begin{gather}\label{eq:aux12}
\frac{\pd}{\pd \zeta}\left(N\sqrt{\eta}\right) + \frac{\pd}{\pd \proj{x}^i}\left(\alpha N\proj{U}^i\sqrt{\eta} \right) = 0,
\end{gather}
where we introduced the flow velocity, as measured by FO, according to $\vec{U} = \favr{\vec{V}}$. Using the expression for the 3D divergence operator written in the metric $\eta_{ij}$ (see, \eg Sec.~4.7 in \Ref{book:weinberg}), one can rewrite \Eq{eq:aux12} as
\begin{gather}\label{eq:conteq}
\frac{1}{\sqrt{\eta}}\,\frac{\pd}{\pd \zeta}\left(N\sqrt{\eta}\right) + \vec{\del} \cdot \left(\alpha N\vec{U}\right) = 0.
\end{gather}
This represents the continuity equation, which could also be obtained from the particle conservation in the 4D spacetime, by requiring that the four-divergence of the particle flow be zero \cite{ref:thorne82}.

As our next step, let us multiply \Eq{eq:divform1} by $\proj{p}^j$ and then integrate over $d^3\proj{p}^*$ again. In this case, one obtains
\begin{multline}\label{eq:hp1}
\frac{1}{\sqrt{\eta}}\,\frac{\pd}{\pd \zeta}\left(N \proj{\mc{P}}^j\sqrt{\eta}\right)
+ \frac{1}{\sqrt{\eta}}\,\frac{\pd}{\pd \proj{x}^i}\left(\alpha N \proj{U}^i \proj{\mc{P}}^j \sqrt{\eta}\right)\\
 + \alpha N \proj{\Gamma}^j_{k\ell}\proj{U}^k\proj{\mc{P}}^\ell+ [\vec{\del} \cdot (\alpha \oper{\Pi})]^j 
= \alpha N(\vec{\mc{F}}+\oper{K} \cdot \vec{\mc{P}})^j.
\end{multline}
Here we introduced the average momentum $\vec{\mc{P}} = \favr{\vec{p}}$, the average force on a particle
\begin{gather}
\vec{\mc{F}} = m \favr{\gamma} \vecgrav + \oper{K} \cdot \vec{\mc{P}} +  q\left(\vec{E} + \vec{U} \times \vec{B}\right),
\end{gather}
and the pressure tensor
\begin{gather}
\proj{\Pi}^{jk} = \int (\proj{p}^j - \proj{\mc{P}}^j)(\proj{V}^k - \proj{U}^k)\,f\,d\mc{V}_{\vec{p}},
\end{gather}
so one can interpret the fourth term in \Eq{eq:hp1} as $j$th component of its divergence (Sec.~4.7 in \Ref{book:weinberg}):
\begin{gather}
[\vec{\del} \cdot (\alpha \oper{\Pi})]^j 
= \frac{1}{\sqrt{\eta}}\,\frac{\pd}{\pd \proj{x}^k}\left(\alpha \proj{\Pi}^{jk} \sqrt{\eta}\right)
+ \proj{\Gamma}^j_{k \ell} \proj{\Pi}^{\ell k}.
\end{gather}
With \Eq{eq:conteq} taken into account, \Eq{eq:hp1} rewrites as
\begin{multline}
\frac{1}{\alpha}\,\frac{\pd \proj{\mc{P}}^j}{\pd \zeta} 
- (\oper{K} \cdot \vec{\mc{P}})^j 
+ (\vec{U}\cdot \vec{\del}) \proj{\mc{P}}^j 
+ \proj{\Gamma}^j_{k\ell}\proj{U}^k\proj{\mc{P}}^\ell  \\
=  - \frac{1}{\alpha N}\,[\vec{\del} \cdot (\alpha \oper{\Pi})]^j + \proj{\mc{F}}^j.
\end{multline}
Then, following the same argument as in \Sec{sec:vectormotion}, one obtains the vector equation
\begin{multline}\label{eq:hp2}
\left[\frac{1}{\alpha}\,\frac{\pd}{\pd \zeta} + (\vec{U}\cdot \vec{\del}) \right] \vec{\mc{P}} 
= - \frac{1}{\alpha N}\,\vec{\del} \cdot (\alpha \oper{\Pi})\\
+ m \favr{\gamma} \vecgrav + \oper{K} \cdot \vec{\mc{P}} +  q\left(\vec{E} + \vec{U} \times \vec{B}\right),
\end{multline}
similar to that in the Minkowski metric.

Higher moments of the Vlasov equation, which could yield a hydrodynamic closure like in \Refs{ref:tokatly99, tex:oberman60, ref:bernstein60, my:dense}, can be obtained analogously, and most easily for a nonrelativistic motion ($\gamma \approx 1$). Those are not discussed here, but are addressed separately in our \Ref{my:mquanta}, where we contemplate the evolution of linear waves in a metric with nonzero $\oper{K}$. Examples showing how the above equations can be used for applied calculations are also given in \Ref{my:mquanta}.

\section{Conclusions} 
\label{sec:conclusions}

In this paper, the 3D dynamics of a charged particle in an arbitrary spacetime metric $g_{\mu\nu}$, traditionally addressed within differential geometry, is reformulated in terms of linear algebra (\Sec{sec:slicing}) and Hamiltonian formalism (\Sec{sec:motion}). The modification of the Vlasov equation, in its standard form describing a charged particle distribution in the 6D phase space, is then derived explicitly, in two equivalent forms [\Eqs{eq:vl1} and \eq{eq:divform1}]. The equation accounts simultaneously for the Lorentz force and the effects of general relativity, with the latter appearing as the gravity force and an additional force due to the extrinsic curvature $\oper{K}$ of spatial hypersurfaces $\Sigma_t$. For an arbitrary spatial metric, the equations of collisionless hydrodynamics are also obtained in the usual three-vector form [\Eqs{eq:conteq} and \eq{eq:hp2}]; for their applications, see our \Ref{my:mquanta}. Another form of the Vlasov equation, which does not lead to vector equations but, on the other hand, allows for an arbitrary spacetime basis $\mc{B}_0$, is also derived [\Eqs{eq:vl2} and \eq{eq:divform2}] within the new formalism and agrees with the results reported in \Refs{ref:debbasch09a, ref:debbasch09b}.

\section{Acknowledgments}

The authors thank A.~I. Zhmoginov for valuable suggestions and critical comments. The work was supported by the NNSA under the SSAA Program through DOE Research Grant No.~DE-FG52-08NA28553.


\gap
\appendix

\section{Another form of the particle motion equation}
\label{app:deriv}

Although \Eq{eq:vec1} is sufficient for our purposes, let us explain how our notation relates to that from the widely cited \Ref{ref:thorne82} (see also \Refs{ref:jantzen92, ref:bini97, tex:bini98a, tex:bini98b}). First of all, the derivative on the left-hand side of \Eq{eq:vec1} is the same as (the spatial part of) the four-vector
\begin{gather}
\foper{h}\cdot \frac{d(\foper{h} \cdot \fvec{p})}{d\taufo} 
 = \gamma^{-1}\foper{h}\cdot \frac{d(\foper{h} \cdot \fvec{p})}{d\tau},
\end{gather}
where $\tau$ is the particle proper time, $d\tau = d\taufo/\gamma$; thus, \Eq{eq:vec1} coincides with Eq.~(3.8) in \Ref{ref:thorne82}. Following \Ref{ref:thorne82}, expand this derivative as
\begin{gather}
\frac{d\vec{p}}{d\taufo} = D_\tau \vec{p} + \frac{1}{\alpha}\, (\vec{v} \cdot \vec{\del}) \vec{p} = D_\tau \vec{p} + (\vec{V} \cdot \vec{\del}) \vec{p}.
\end{gather}
Here $D_\tau$ is called the Fermi-Walker derivative,
\begin{gather}
D_\tau \vec{p} \equiv \frac{1}{\alpha}\, \frac{\pd \vec{p}}{\pd \zeta}.
\end{gather}
which applies to \textit{both} the components $\proj{p}^i$ and the basis vectors $\vec{e}_i$ that constitute the three-vector $\vec{p} = \vec{e}_i \proj{p}^i$. (The index $\tau$ denotes differentiating with respect to $\taufo$ rather than $\tau$.) One can further rewrite $D_\tau \vec{p}$ as
\begin{gather}
D_\tau \vec{p} = \frac{1}{\alpha}\left(\vec{e}_i \,\frac{\pd \proj{p}^i}{\pd \zeta} + \frac{\pd \vec{e}_i}{\pd \zeta}\, \proj{p}^i\right).  
\end{gather}
Hence, its projection on the $\ell$th axis equals
\begin{gather}
(D_\tau \vec{p})^\ell = \frac{1}{\alpha}\,\frac{\pd \proj{p}^\ell}{\pd \zeta} - \proj{K}^\ell_k \proj{p}^k, 
\end{gather}
where we used \Eq{eq:kij}. Reverting to the vector form, one can write then
\begin{gather}
D_\tau \vec{p} = \mc{D}_\tau \vec{p} - \oper{K} \cdot \vec{p}
\end{gather}
(cf. Eq.~(2.16b) in \Ref{ref:thorne82}), where, by definition,
\begin{gather}\label{eq:mcd1}
\mc{D}_\tau \vec{p} \equiv \mc{D}_\tau (\vec{e}_\ell \proj{p}^\ell) 
= \frac{\vec{e}_\ell}{\alpha}\,\frac{\pd \proj{p}^\ell}{\pd \zeta}.
\end{gather}
That is, the new derivative $\mc{D}_\tau$ acts on the vector components but does not affect $\vec{e}_i$. Notice now that, in the basis $\mc{B}_\zeta$, the partial derivatives of vector components with respect to $\zeta$ coincide with the Lie derivatives along $\upzeta_* \equiv \alpha \fvec{n}$ (see, \eg \Ref{arX:gourgoulhon07}). Since this holds for all components of $\mc{D}_\tau \vec{p}$, the latter can be given the following covariant definition:
\begin{gather}\label{eq:mcd2}
\mc{D}_\tau \vec{p} = \frac{\vec{e}_\ell}{\alpha}\,(\lie_{\alpha \fvec{n}} \fvecp{p})^\ell,
\end{gather}
that is, $\alpha\mc{D}_\tau$ can be understood as (the spatial part of) the Lie derivative along $\alpha \fvec{n}$ (cf. Eq.~(2.13) in \Ref{ref:thorne82}).

Similarly, one can introduce yet another Lie derivative, now along the original time axis $\fvec{e}_0$ (rather than $\upzeta_*$). Using \Eq{eq:e0gen}, one obtains
\begin{gather}
\lie_t \equiv  \lie_{\fvec{e}_0} = \lie_{\alpha \fvec{n}} + \lie_\upbeta.
\end{gather}
Thus, for three-vectors,
\begin{gather}
\lie_t \vec{p} = \alpha\,\mc{D}_\tau \vec{p} + \lie_\upbeta \vec{p}
\end{gather}
(cf. Eq.~(2.16c) in \Ref{ref:thorne82}). Again use that, in the basis $\mc{B}_0$, the Lie derivative along $\fvec{e}_0$ coincides with the partial derivative with respect to $t$: $(\lie_t \fvecp{p})^\ell = \pd \proj{p}^\ell/\pd t$. Then,
\begin{gather}
\lie_t \fvecp{p}  = \fvec{e}_\ell (\lie_t \fvecp{p})^\ell = \fvec{e}_\ell\,\frac{\pd \proj{p}^\ell}{\pd t}.
\end{gather}
In other words, $\lie_t$ can be understood as the time derivative, which differentiates the vector components but not the basis vectors in $\mc{B}_0$.

Combining the above formulas, one gets
\begin{gather}
D_\tau = \alpha^{-1}(\lie_t  - \lie_\upbeta) - \oper{K}.
\end{gather}
Hence, the particle motion equation, \Eq{eq:vec1}, rewrites equivalently as
\begin{multline}\label{eq:vec2}
\alpha^{-1}\left[\lie_t + (\vec{v} \cdot \vec{\del}) \right]  \vec{p}
 =  m \gamma \vecgrav + 2 \oper{K} \cdot \vec{p} + \alpha^{-1}\lie_\upbeta\vec{p} \\ + q\left(\vec{E} + \vec{V} \times \vec{B}\right),
\end{multline}
where $\oper{K}$ can be alternatively put as \cite{ref:thorne82}
\begin{gather}\label{eq:ksigma}
\oper{K} = - \oper{\sigma} - (\theta/3)\,\oper{I},
\end{gather}
with $\oper{\sigma}$ being the traceless ``shear tensor'', $\theta = \text{Tr}\, \oper{K}$ being the volume expansion rate, and $\oper{I}$ being the unit tensor. (Also, see \Ref{ref:thorne82} and Sec.~21.5 in \Ref{book:misner77} for the relation between $\oper{K}$ and $\fvec{n}(x^\mu)$.) This allows a direct comparison of our result with that given in Eq.~(17) in the widely cited \Ref{ref:holcomb89}. Particularly, one can see that the latter must be corrected in the following aspects: (i)~the time derivative that enters \Eq{eq:vec2} is, strictly speaking, the Lie derivative $\lie_t$; (ii)~ the velocity $\vec{v}$ on the left-hand side is different from $\vec{V}$ on the right-hand side [\Eq{eq:V}]; (iii)~the Lorentz factor $\gamma$ is determined by $\vec{V}$ but not $\vec{v}$ [\Eq{eq:gammaV}]; (iv)~the gravitational force $m \gamma \vecgrav$ enters the right-hand side with the plus sign rather than the minus sign.


\begin{thebibliography}{10}

\bibitem{tex:ehlers71}
J. Ehlers, \textit{General relativity and kinetic theory}, in \textit{General
  Relativity and Cosmology} (Academic Press, New York, 1971), edited by B.~K.
  Sachs, p.~1.

\bibitem{book:cercignani}
C.~Cercignani and G.~M. Kremer, {\it The Relativistic Boltzmann Equation:
  Theory and Applications\/} (Birkh\"auser, Boston, 2002).

\bibitem{ref:thorne82}
K.~S. Thorne and D.~Macdonald, Mon. Not. R. Astr. Soc. {\bf 198}, 339 (1982).

\bibitem{ref:evans88}
C.~R. Evans and J.~F. Hawley, Astrophys. J. {\bf 332}, 659 (1988).

\bibitem{arX:gourgoulhon07}
\'E. Gourgoulhon, arXiv:gr-qc/0703035 (2007).

\bibitem{ref:jantzen92}
R.~T. Jantzen, P.~Carini, and D.~Bini, Ann. Phys. {\bf 215}, 1 (1992).

\bibitem{ref:bini01}
D.~Bini, C.~Germani, and R.~T. Jantzen, Intl. J. Mod. Phys. D {\bf 10}, 633
  (2001).

\bibitem{ref:bini97}
D.~Bini, P.~Carini, and R.~T. Jantzen, Intl. J. Mod. Phys. D {\bf 6}, 143
  (1997).

\bibitem{tex:bini98a}
D. Bini, P. Carini, R.~T. Jantzen, \textit{The inertial forces/Test particle
  motion game}, in \textit{Proceedings of the 8th Marcel Grossmann Meeting on
  General Relativity} (World Scientific, Singapore, 1998), edited by T. Piran,
  p.~376 [arXiv:gr-qc/9710051 (1997)].

\bibitem{tex:bini98b}
D. Bini, F. de Felice, R.~T. Jantzen, \textit{Centripetal acceleration and
  centrifugal force in general relativity}, in \textit{Nonlinear
  Gravitodynamics: The Lense-Thirring effect}, edited by R. Ruffini and C.
  Sigismondi (World Scientific, Singapore, 2003), p.~119 [Proceedings of the
  First ICRA Network Workshop on the Lense-Thirring Effect (1998)].

\bibitem{book:stix}
T.~H. Stix, {\it Waves in Plasmas\/} (AIP, New York, 1992).

\bibitem{book:bernstein}
J.~Bernstein, {\it Kinetic Theory in the Expanding Universe\/} (Cambridge
  University Press, New York, 1988).

\bibitem{ref:dettmann93}
C.~P. Dettmann, N.~E. Frankel, and V.~Kowalenko, Phys. Rev. D {\bf 48}, 5655
  (1993).

\bibitem{ref:debbasch09a}
F.~Debbasch and W.~A. van Leeuwen, Physica A {\bf 388}, 1079 (2009).

\bibitem{ref:debbasch09b}
F.~Debbasch and W.~A. van Leeuwen, Physica A {\bf 388}, 1818 (2009).

\bibitem{foot:correct}
For example, see Appendix for comparison between \Ref{ref:thorne82} and \Ref{ref:holcomb89}, both of which are widely cited.

\bibitem{ref:gailis94}
R.~M. Gailis, C.~P. Dettmann, N.~E. Frankel, and V.~Kowalenko, Phys. Rev. D
  {\bf 50}, 3847 (1994).

\bibitem{ref:gailis95}
R.~M. Gailis, N.~E. Frankel, and C.~P. Dettmann, Phys. Rev. D {\bf 52}, 6901
  (1995).

\bibitem{ref:gailis97}
R.~M. Gailis and N.~E. Frankel, Phys. Rev. D {\bf 56}, 7750 (1997).

\bibitem{ref:zhang89}
X.-H. Zhang, Phys. Rev. D {\bf 39}, 2933 (1989).

\bibitem{ref:elst97}
H.~van Elst and C.~Uggla, Class. Quantum Grav. {\bf 14}, 2673 (1997).

\bibitem{ref:baumgarte03}
T.~W. Baumgarte and S.~L. Shapiro, Astrophys. J. {\bf 585}, 921 (2003).

\bibitem{ref:sakai03}
N.~Sakai and S.~Shibata, Astrophys. J. {\bf 584}, 427 (2003).

\bibitem{ref:katanaev06}
M.~O. Katanaev, Theor. Math. Phys. {\bf 148}, 1264 (2006).

\bibitem{ref:zanna07}
L.~Del Zanna, O.~Zanotti, N.~Bucciantini, and P.~Londrillo, Astron. Astrophys.
  {\bf 474}, 11 (2007).

\bibitem{foot:dual}
We deliberately avoid introducing the dual space and differential forms;
  otherwise see, \eg Chap.~8 in \Ref{book:misner77}.

\bibitem{book:misner77}
C.~W. Misner, K.~S. Thorne, and J.~A. Wheeler, {\it Gravitation\/} (Freeman, S.
  Francisco, 1973).

\bibitem{book:weinberg}
S.~Weinberg, {\it Gravitation and Cosmology: Principles and Applications of the
  General Theory of Relativity\/} (Wiley, New York, 1972).

\bibitem{book:landau2}
L.~D. Landau and E.~M. Lifshitz, {\it The Classical Theory of Fields\/}
  (Pergamon Press, New York, 1971).

\bibitem{ref:cognola86}
G.~Cognola, L.~Vanzo, and S.~Zerbini, Gen. Relativ. Gravit. {\bf 18}, 971
  (1986).

\bibitem{ref:dewitt66}
B.~S. DeWitt, Phys. Rev. Lett. {\bf 16}, 1092 (1966).

\bibitem{ref:holcomb89}
K.~A. Holcomb and T.~Tajima, Phys. Rev. D {\bf 40}, 3809 (1989).

\bibitem{foot:liecurv}
$\pd/\pd \zeta$ can be understood as the Lie derivative along $\upzeta_*$; see
  also Appendix.

\bibitem{book:landau10}
E.~M. Lifshitz and L.~P. Pitaevskii, {\it Physical Kinetics\/} (Pergamon Press,
  New York, 1981).

\bibitem{foot:dform}
Strictly speaking, $d^3\proj{x}^i$ and $d^3\proj{p}_i$ are three-forms
  \cite{ref:debbasch09a}.

\bibitem{foot:sign}
The sign difference in the definition of $r$ from \Ref{ref:debbasch09a} is due
  to the difference in the metric signatures.

\bibitem{ref:tokatly99}
I.~Tokatly and O.~Pankratov, Phys. Rev. B {\bf 60}, 15550 (1999).

\bibitem{tex:oberman60}
C. Oberman, {\it On the correspondence between solutions of the collisionless
  equation and the derived moment equations}, PPPL tech. report Matt-57 (1960).

\bibitem{ref:bernstein60}
I.~B. Bernstein and S.~K. Trehan, Nuclear Fusion {\bf 1}, 3 (1960).

\bibitem{my:dense}
I.~Y. Dodin, V.~I. Geyko, and N.~J. Fisch, Phys. Plasmas {\bf 16}, 112101
  (2009).

\bibitem{my:mquanta}
I.~Y. Dodin and N.~J. Fisch, Phys. Rev. D {\bf 82}, 044044 (2010).

\end{thebibliography}

\end{document}